\documentclass[twocolumn, pra]{revtex4-1}

\usepackage{amsmath}
\usepackage{graphicx}
\usepackage{amsfonts}
\usepackage{amsthm}
\usepackage{enumitem}
	\newtheorem{theorem}{Theorem}
	\newtheorem{defi}{Definition}
\usepackage[usenames,dvipsnames]{color}
\usepackage[colorlinks=true, citecolor=blue,urlcolor=blue, linkcolor=blue]{hyperref}

\makeindex

\begin{document}
\title{Exact dimension estimation of interacting qubit systems assisted by a single quantum probe}

\author{Akira Sone}
\author{Paola Cappellaro}
\email{pcappell@mit.edu}
\affiliation{Research Laboratory of Electronics and Department of Nuclear Science and Engineering, Massachusetts Institute of Technology, Cambridge, MA 02139}
\begin{abstract}
Estimating the dimension of an Hilbert space is an important component of quantum system identification. 
In quantum technologies, the dimension of a quantum system (or its corresponding accessible Hilbert space) is an important resource, as larger dimensions determine e.g. the performance of quantum computation protocols or the sensitivity of quantum sensors. 
{Despite} being a critical task in quantum system identification, estimating the Hilbert space dimension is experimentally challenging. While there have been proposals for various dimension witnesses capable of putting a lower bound on the dimension from measuring collective observables that encode correlations, in many practical scenarios, especially for multiqubit systems, the experimental control might not be able to engineer  the required initialization, dynamics and observables. 

Here we propose a more practical strategy that relies not on directly measuring an unknown multiqubit target system, but on the indirect interaction with a local quantum probe under the experimenter's control.  
Assuming only that the interaction model is given {and the evolution correlates all the qubits with the probe}, we combine a graph-theoretical approach and realization theory to demonstrate that the system dimension can be \textit{exactly} estimated from the model order of the system. 
We further analyze the robustness in the presence of background noise of the proposed estimation method based on realization theory, finding that despite stringent constrains on the allowed noise level,  exact dimension estimation can still be achieved.
\end{abstract}
\maketitle

\section{Introduction}
\label{sec:intro}
The dynamics of a closed quantum system is determined by its Hamiltonian thus its identification is a central task for all quantum protocols. In finite interacting qubit systems, the Hamiltonian can be usually characterized by single qubit energies and  qubit-qubit interactions, and, importantly, by the system dimension, that is, the number of qubits. 
These three sets of parameters include all the information describing the  system properties. For example, in a spin-1/2 system, identifying Zeeman energy shifts~\cite{Zhang14, Matsuzaki16a, Wang16} yields information on the spin species. By identifying the spin-spin interaction Hamiltonian, we  obtain  information on (1) the system graph~\cite{Kato14}, (2) the {coupling type},  and (3) the relative spin positions from the coupling strengths~\cite{Burgarth09b, DiFranco09, Zhang14, Wang16}. Identifying only the first two pieces of information (graph and coupling type) enables writing a general model for the spin system. Then, to further specify the system, one needs to identify not only the coupling strengths, but also importantly the number of spins in the system.
In this paper, we  focus on  estimating the dimension of the Hilbert space (or the number of qubits) under the assumption that we know (1) the graph structure of the system and (2) the coupling type.

The dimension of the Hilbert space (or system dimension) is indeed an important information for any quantum device. The performance of quantum protocols, such as the computational complexity of quantum algorithms~\cite{Nielsen00b} or quantum process tomography~\cite{Wang16}, is strictly dependent on the dimension. In addition, in our recent work~\cite{Sone17a}, we demonstrated that the dimension also determines what experimental resources, such as the number of sampling points and the total time evolution, are needed to characterize the rest of the Hamiltonian parameters. These dimension-dependent quantities are important for practical applications of quantum engineering. Therefore, dimension estimation is a significant task in quantum system identification.

The estimation of the system dimension was first addressed by Brunner \textit{et al}~\cite{Brunner08} by introducing the concept of dimension witnesses. Dimension witnesses are a test giving a \textit{lower bound} on the system dimension that can reproduce the measurement data~\cite{Brunner08, Wehner08, Vertesi09, Wolf09, Gallego10, Brunner13, Guhne14, Bowles15, Sikora15, Vicente17}. Recently, there have been  experimental efforts to demonstrate several theoretical proposals for  dimension witnesses~\cite{Hendrych12, Ahrens14, Cai16} that typically require measuring all correlations generated in the system.

Here, we are interested instead in \textit{exact} dimension estimation.
In  previous studies, a lower bound on the dimension was provided just from performing  correlation measurements in an arbitrary quantum system, without any prior information about the Hamiltonian. Here we assume instead that the interaction model (defined more precisely in Def.~\ref{defi:model}) is given as prior information. However, although graph structure and coupling type are known, the exact Hamiltonian is still unknown, since we cannot hope to measure all the correlations in the system if the dimension of the target system is unknown. 
We thus focus on an alternative, and more practical scenario, where the target system cannot be directly observed or controlled, but interacts with a single \textit{quantum probe} under the experimentalist's control. We will exploit the dynamics of the quantum probe to estimate the target system's dimension.

To identify a linear time-invariant (LTI) system {in the state-space representation}, a popular system identification methodology~\cite{Katayama05} is the eigensystem realization algorithm (ERA), which is derived from  realization theory. Zhang and Sarvoar~\cite{Zhang14} discussed the first application of ERA to quantum  Hamiltonian parameter estimation. A key step of the algorithm is the singular value decomposition of a Hankel matrix, whose elements are the measurement data at equally spaced {sampling times}. 
{In the noiseless case, the rank of the Hankel matrix is equivalent to the model order, which is the {degree of the characteristic polynomial of the irreducible transfer function} describing the system dynamics in the Laplace space~\cite{Antsaklis07}.} 
{This can be interpreted as the minimum number of independent state variables required to fully describe the dynamics of the system.}  

In a general many-body interacting system, the state variables are the observables, including the observable to be directly measured and the operators generated from the dynamics, which can be \textit{indirectly} observed and controlled. Then, the key insight into dimension estimation is that the number of generated correlations will strongly depend on the system dimension, and we can thus expect that the dimension will be a function of the model order, which is revealed by {realization theory} and experimental measurements.

The paper is organized as follow. {We first provide an intuitive description of the  dimension estimation scheme in Sec.~\ref{sec:SimpSche}, in order to provide a simple example of what would be necessary to implement it in practice. This simple example, and in particular the goal of the dimension estimation scheme and the prior information needed, is made more precise in Sec.~\ref{sec:preli}, where we provide the definition of the interaction model by applying recursively constructible families of graphs.} Then, in Sec.~\ref{sec:noiseless} we demonstrate that the exact dimension estimation of the system can be achieved via single-probe measurement if the interaction model is given as prior information and all the qubits are correlated to the probe. As an exemplary system, we consider the one-dimensional spin chain model with nearest-neighbor coupling. We also discuss the estimation performance in the presence of noise in Sec.~\ref{sec:noisy}, before concluding remarks in Sec.~\ref{sec:conc}.

\section{Intuitive picture of dimension estimation scheme}\label{sec:SimpSche}
Before giving the details of the dimension estimation protocol, we present a simple example in order to present a more concrete picture of the scheme.
We assume that the target system can be characterized by a known \textit{interaction model} (see Section~\ref{sec:preli}). For example, we could  consider a spin-1/2 chain with XY Hamiltonian,
\begin{align*}
  H=\sum_{k=1}^{N-1} \frac{J_k}
  {2} (S_k^xS_{k+1}^{x}+S_k^yS_{k+1}^y) 
\end{align*}
(we will use this same example in Sec.~\ref{sec:noisy} to demonstrate dimension estimation in the presence of noise.)
Although we know the general model of the system, the Hamiltonian is still unknown since we do not have the values $J_k$ and, importantly, we do not know the chain length $N$. The goal is to estimate $N$, {so that we can obtain the system dimension $2^N$}, by letting the target spin chain interact with a quantum probe (here spin \#1).

The dynamics of the quantum probe is measured at successive time steps by measuring the expectation value of a (set of) quantum observable(s) $O(t)$ on the probe. To do so, we repeatedly initialize the probe in a state such that $\langle O(0)\rangle\neq 0$, and let the system evolve for time $t_j=jdt$, $j=0,1,\cdots,z$, acquiring the expectation values $\{y(j)\}=\{\langle O(t_j)\rangle\}$. These measurement results can be used in eigenvalue realization theory to determine the size of the system (with a procedure explained in details in Section~\ref{sec:noiseless}), by considering an increasing number $z$ of total measurements. Intuitively, this is because if the number of measurements is too small, the set of  $\{y(j)\}$ is not enough to identify the system; but after measuring enough $y(j)$, additional measurement do not provide additional information, and this is reflected in the properties of the \textit{realization} built from the measurement outcomes. Finally, the prior information about the interaction model can be used to convert from the realization properties to the system dimension. 

In the following two sections we define more rigorously the interaction model and provide an explicit procedure for determining the system dimension from realization theory.

\section{Interaction Model}
\label{sec:preli}
Our dimension estimation method requires some prior information on the target quantum system. Since the system Hamiltonian and dimension are unknown, we want a prescription for carefully defining what prior information is needed, which relies on graph theory.

Generally, an interacting qubit system can be represented by a graph $\mathcal{G}=(V(\mathcal{G}),E(\mathcal{G}))$ with the set of vertices $V(\mathcal{G})$ representing qubits and the set of edges $E(\mathcal{G})$ representing the connectivity between every pair of qubits~\cite{Christandl04}. In the following, we denote by $|V(\mathcal{G})|$ the number of vertices, and by $|E(\mathcal{G})|$ the  number of  edges in the graph $\mathcal{G}$. Operations on the graph are called  \textit{primary operations}, and include adding and deleting vertices and edges~\cite{Gross13}. While edges generally describe qubit-qubit interactions, here we keep the network topology separated from the actual coupling strength, with the graph only encoding information on the former. Thus, we consider an unweighted and undirected graph. We characterize the graph of the system by taking the following rules on edges:
\begin{enumerate}[leftmargin=3pt,itemindent=*, label={\bfseries Rule \arabic*:}]
\item The {single-qubit} energy term can be described by a \textit{self-loop}, which joins a single vertex to itself.
\item The qubit-qubit interaction term can be described by a \textit{proper edge}, which joins two distinct vertices. 
\end{enumerate}
Taking into account these rules, an interacting qubit system can be described by the adjacency matrix $\Xi(\mathcal{G})$:
\begin{align*}
[\Xi(\mathcal{G})]_{ij}=
\begin{cases}
1,~\text{a vertex $v_{i}$ has self-loop, i.e. $i=j$.}\\ 
1,~\text{vertices $v_{i}$ and $v_{j}$ are adjacent.}&\\
0,~\text{otherwise}&,
\end{cases}
\end{align*}
which shows the graph structure of the system.

The interaction model is determined by the graph structure and the {coupling type} between every pair of qubits. Before introducing our definition of interaction model, let us first explain the logic flow to obtain the definition. Recall that we are interested in the relation between the dimension and the model order. 
{Let us denote by {$\text{dim}(\mathcal{H})=2^N$} the dimension of the Hilbert space $\mathcal{H}$ and by $n$ the model order. Considering the $N$-qubit system, we are interested in the function $f:\mathbb{N}\to\mathbb{N}$ such that $N=f(n)$. Although we cannot always analytically derive the exact form of $f$, we are still able to derive a \textit{recurrence relation} that $f$ must satisfy. Once we obtain $f$, then we can obtain the system dimension as a function of the model order.}

As mentioned, the graph structure is determined by the adjacency matrix. When we say ``fixing the graph structure'', we mean that for a given sequence of graphs $\{\mathcal{G}_{k}\}_{k\ge1}$, the rule that each corresponding adjacency matrix satisfies is fixed. Recall that the construction of the adjacency matrix is dependent only on the elementary operations. 
Therefore, fixing the rule means that a sequence of graphs $\{\mathcal{G}_{k}\}_{k\ge1}$ is constructed from a given initial graph $\mathcal{G}_{0}$ by a repeating a succession of fixed elementary operations. 
In  graph theory, $\{\mathcal{G}_{k}\}_{k\ge0}$ is called \text{recursively constructible family of graphs}~\cite{Noy04}. 
For example, let us consider the following one-dimensional Ising model without transverse fields, i.e. no self-loops: $H=\sum_{k=1}^{N-1}\frac{J_k}{2} S^{\alpha}_{k}S^{\alpha}_{k+1}$, where $iS^{\alpha}_{k}\in \mathfrak{su}(2)$ \cite{Hall15}, is an arbitrary operator acting on $k$-th spin. 
The fixed elementary operation is to add one vertex and proper edge from the right side which joins neighboring vertices. Let us label each spin by $1,2,3,\cdots$ from the left side. In this case, the initial graph $\mathcal{G}_{0}$ is the first spin. The corresponding adjacency matrices for the sequence of graphs $\{\mathcal{G}_1, \mathcal{G}_2, \mathcal{G}_{3},\cdots\}$ are given by:
\begin{align*}
[\Xi(\{\mathcal{G}_{k\ge 1}\})]_{i,j}=
\begin{cases}
1,~\text{when $i=j\pm1$}\\
0,~\text{otherwise}.
\end{cases}
\end{align*}

\begin{figure}[tbh]
\includegraphics[width=0.48\textwidth]{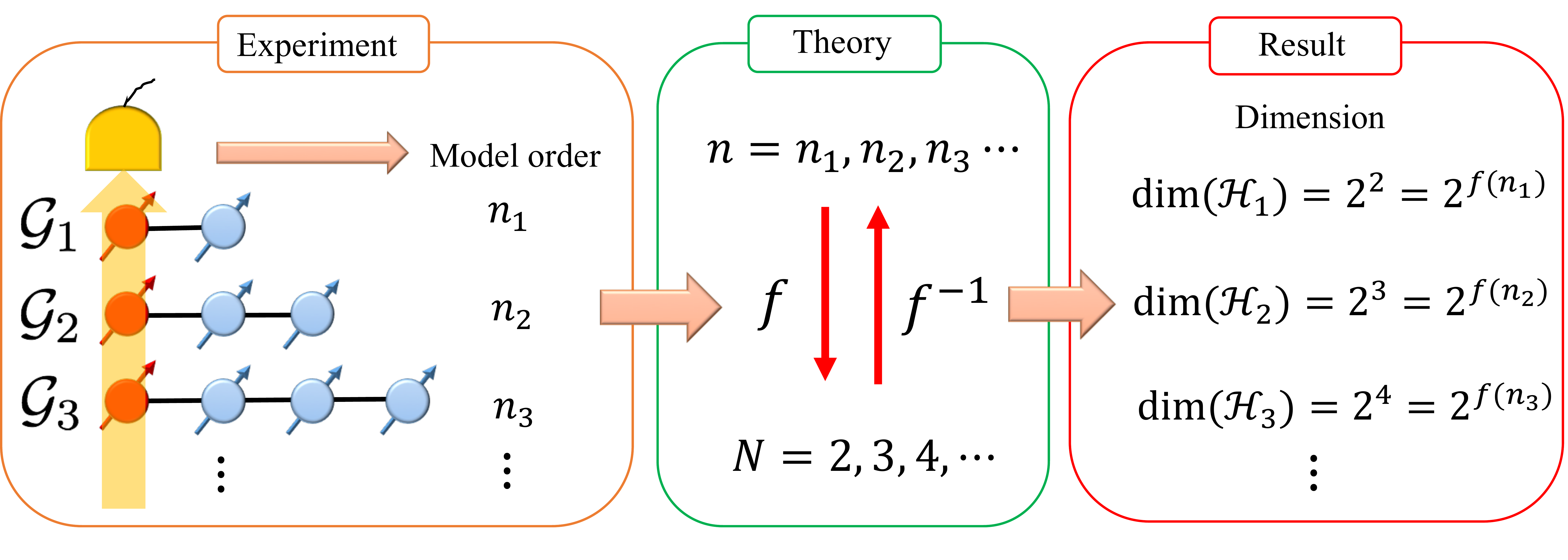}
\caption{\textbf{Dimension estimation conceptual scheme.} We first derive the function $f$ such that {$N=f(n)$}. Through the measurement, we can determine the model order $n$, yielding the system dimension as $\text{dim}(\mathcal{H})=2^N=2^{f(n)}$.}
\label{fig:DimEst}
\end{figure}

Suppose that we want to estimate the dimension by measuring only the first spin (quantum probe). We first have to obtain the relation $N=f(n)$, which depends on the coupling type between every pair of spins and the rule that each adjacency matrix has to satisfy. Then, we find the model order $n$ through our measurement to obtain the dimension $\text{dim}(\mathcal{H})=2^{f(n)}$ (See Fig.~\ref{fig:DimEst}). 

Since elementary operations can be defined in any spatial dimensions, this approach is valid generally even for higher spatial dimensions. Now, we are ready to define an interaction model. Taking into account the fact that we are interested only in the connectivity of the spins, the graphs must be connected, undirected, and unweighted. Then, an interaction model can be defined as the following:
\begin{defi}
\textbf{Interaction model}: An interaction model $\mathcal{M}$ is defined by a coupling type and a recursively constructible family of connected graphs $\{\mathcal{G}_{k}\}_{k\ge0}$, which are undirected and unweighted. 
\label{defi:model}
\end{defi}

In the following, when we mention fixing an interaction model, we always mean that we fix (1) the coupling type, (2) the initial graph $\mathcal{G}_{0}$, and (3) the repeated succession of elementary operations. These steps lead the sequence of $\{|V(\mathcal{G}_{k})|\}_{k\ge0}$ to satisfy an explicit recurrence relation. In the following, we clarify this point by combining the graph-theoretic approach just discussed, and the ERA approach to Hamiltonian identification~\cite{Zhang14}.

\section{Noiseless dimension estimation}
\label{sec:noiseless}
As explained in Sec.~\ref{sec:SimpSche}, we wish to estimate the system dimension by looking at the properties (in particular the model order) of its realization, which can be evaluated by experimental measurements. To achieve this goal, we want to relate the dimension $2^N$ of the qubit system to the model order $n$. 
We proceed by first relating the system dimension to the number of the elements in the \textit{accessible set}, and then linking it to the model order.

\subsection{Accessible set}
We consider an {$N$-qubit system} which can be represented using an orthonormal operator basis $B=\{iO_i\}$. As explained later, the chosen basis must yield a minimal state-space realization of the quantum system. In our examples, this condition is obtained for a particularly simple choice, where the operators $O_i$ are tensor products of Pauli matrices and identity matrix, such that $iO_i \in \mathfrak{su}(2^N)$. 

For an interacting $N$-qubit system, the Hamiltonian $H$ can be written as:
\begin{align*}
H=\sum_{m=1}^{M}\theta_m S_m,
\end{align*}
where $\theta_m\in\mathbb{C}\setminus\{0\}$ are the parameters, and $iS_m\in B$. 
The set $\Gamma$ of these operators, 
\begin{align*}
\Gamma=\{S_m~|~iS_{m}\in B,~m=1,2,\cdots,M\},
\end{align*}
is a subset of the whole basis and  can be interpreted as a \textit{representation} of the interaction model $\mathcal{M}$ for a fixed dimension $2^N$. 
We further introduce the \textit{observable} set 
\begin{align*}
G_0=\{\Pi_l~|~i\Pi_l\in B,~[\Pi_{l},H]\neq 0\}.
\end{align*}
Here we restrict the $\Pi_l$  to be observables on the quantum probe, in accordance with our model, even if more generally the observable  set can include any observable. 
Note that if we consider not a basis element, but a general observable  $\Xi_{p}=\sum_{l}\nu_{l}^{(p)}\Pi_{l}~(\nu_{l}^{(p)}\in\mathbb{R})$,  we should include into the observable set $G_0$ all the basis operators in the linear expansion of  $\Xi_{p}$   (See~\cite{Zhang14} for an example).

For a given graph $\mathcal{G}$, we consider the following iterative procedure \cite{Sastry99, Zhang14, Zhang15, Sone17a}:
\begin{equation}
G_j(\mathcal{G})\equiv G_{j-1}(\mathcal{G})\cup [\![G_{j-1}(\mathcal{G}),\Gamma]\!],
\label{eq:iterative}
\end{equation} 
where 
\begin{align*}
[\![G_{j-1}(\mathcal{G}),\Gamma]\!]\equiv \{O_{i}|\text{tr}(O_{i}^{\dagger}[g,\gamma])\neq 0,~g\in G_{j-1}(\mathcal{G}),\gamma\in \Gamma\},
\end{align*}
(See Appendix~\ref{sec:example1} for an example of the construction.)

Due to the finite dimension of $\mathfrak{su}(2^{N})$, the iterative procedure in Eq.~(\ref{eq:iterative}) will saturate, and we can obtain a saturated set of operators $G(\mathcal{G})$.  $G(\mathcal{G})$ is called \textit{accessible set}, and  includes {all the operators  generated from the observable set through the dynamics, 
in particular those describing correlations between the quantum probe and other qubits in the system. Let us denote the number of elements in the accessible set by $|G|$. 

Note that given a basis $B$, if we fix the observable set $G_0$ and the model representation $\Gamma$, $G(\mathcal{G})$ is uniquely determined and has a finite size; therefore, $|G|$ is also unique. This is true for any dimension of the model representation: for a given model $\mathcal{M}$, {if all the qubits in the system are correlated with the quantum probe}, increasing the graph size  $\mathcal{G}_{k}$ will lead to an increase of the accessible set dimension. Then, for a fixed  basis $B$ 
and observable set $G_0$ (yielding a $G$ containing correlations with all the system's qubits), a succession of elementary operations on the graph leads to a recurrence relation satisfied by a sequence $\{|G(\mathcal{G}_{k})|\}_{k\ge 0}$. As the increase in $|V(\mathcal{G}_{k})|$ leads to an increase in $|G(\mathcal{G}_{k})|$, $|G(\mathcal{G})|$ is a strictly increasing function of $N$, 
\begin{equation}
\frac{d}{dN}|G|>0.
\label{eq:dimGN}
\end{equation}
\subsection{Model Order and System Dimension}
While Eq.~(\ref{eq:dimGN}) ensures that one can determine the system dimension (or $N$) from $|G|$, we would like to relate $N$ to a directly measurable quantity, the model order $n$. 
In particular, here we show that
if we choose an operator  basis $B$ such that it yields an equivalent classical system with a \textit{minimal} state-space representation  (that is, both controllable and observable as defined below), we will have $n=|G|$, thus providing the desired relationship between model order and system dimension. 

{First, let us write the state-space representation of the system.} As this approach has already been described elsewhere~\cite{Zhang14, Sone17a}, here we give only a short review as relevant to our dimension estimation scheme. 
  
{Given a basis $B =\{iO_l\}$, } let $x_{l}(t)$ be the expectation value of its elements $O_l$. 
From the basis vectors in $G(\mathcal{G})$, we  construct the  coherent vector, {$\mathbf{x}(t)=(x_1(t),\cdots, x_{|G|}(t))^{T}\in\mathbb{R}^{|G|\times 1}$} {with dimension $\text{dim}(\mathbf{x})=|G|$}. Its time evolution is governed by $\dot{\mathbf{x}}(t)=\tilde{\mathbf{A}}\mathbf{x}(t)$, where {$\tilde{\mathbf{A}}\in\mathbb{R}^{|G|\times |G|}$} is a skew-symmetric matrix, with non-zero elements given by the Hamiltonian parameters. Let $y(t)\in\mathbb{R}$ be the output data obtained via the output matrix {$\mathbf{C}\in\mathbb{R}^{1\times |G|}$}. Then, the dynamics of the system can be described by the following ``classical'' state-space representation:
\begin{equation}
\dot{\mathbf{x}}(t)=\tilde{\mathbf{A}}\,\mathbf{x}(t),~~ y(t)=\mathbf{C}\,\mathbf{x}(t)
\label{eq:linear1}
\end{equation}
The quantum evolution has thus been reduced to an equivalent classical LTI system, with   controllability and observability properties given by the usual classical definition~\cite{Katayama05, Kailath80, Schutter00}.
$(\tilde{\mathbf{A}}, \mathbf{C}, \mathbf{x}(0))$ in Eq.~(\ref{eq:linear1}) is  the \textit{realization} of the irreducible transfer function $T(s)$,
\begin{align*}
T(s)=\mathbf{C}(s\mathbf{I}-\tilde{\mathbf{A}})^{-1}\mathbf{x}(0)=\frac{P(s)}{Q(s)},
\end{align*}
where $\mathbf{I}$ is the $|G|\times |G|$ identity matrix and the system initial state must be chosen such that $\mathbf{x}(0)\neq 0$. $P(s)$ and $Q(s)$ are  polynomials in $s$, and in particular $Q(s)$ is called the characteristic polynomial of $T(s)$. The model order $n$ is defined as \cite{Antsaklis07}: 
\begin{equation}
n=\text{deg}(Q(s)).
\label{eq:degree}  
\end{equation}

From a practical point of view, it is convenient to consider Eq.~(\ref{eq:linear1}) in the discrete-time representation:
\begin{equation}
\mathbf{x}(j+1)=\mathbf{A}\mathbf{x}(j), ~~y(j)=\mathbf{C}\mathbf{x}(j),
\label{eq:discrete}
\end{equation}   
where we set $\mathbf{x}(j)\equiv \mathbf{x}(jdt)$, $y(j)\equiv y(jdt)$ and $\mathbf{A}\equiv e^{\tilde{\mathbf{A}}dt}$.
Since any matrix exponential is a nonsingular matrix, we have: 
\begin{align*}
\text{rank}(\mathbf{A})=\text{dim}(\mathbf{x}(t))=|G|.
\label{eq:rankA}
\end{align*}
From Eq.~(\ref{eq:discrete}), we can construct the following Hankel matrix:
\begin{align*}
\mathbf{H}_{rs}=
\begin{pmatrix}
y(0)&y(1)&\cdots& y(s-1)\\
y(1)&y(2)&\cdots& y(s)\\
\vdots&\vdots&\ddots&\vdots\\
y(r-1)&y(r)&\cdots& y(r+s-2)
\end{pmatrix}.
\end{align*} 
$\mathbf{H}_{rs}$ can be decomposed into
\begin{align*}
\mathbf{H}_{rs}
=\mathcal{O}_{r}\mathcal{C}_{s},
\end{align*}
where $\mathcal{O}_{r}$ and $\mathcal{C}_{s}$ are called \textit{observability} and \textit{controllability} matrix, respectively, defined as:
\begin{align*}
\begin{split}
\mathcal{O}_{r}&=\begin{pmatrix}
\mathbf{C}&
\mathbf{C}\mathbf{A}&
&\cdots
&\mathbf{C}\mathbf{A}^{r-1}
\end{pmatrix}^{T}\\
\mathcal{C}_{s}&=\begin{pmatrix}
\mathbf{x}(0)&\mathbf{A}\mathbf{x}(0)&\cdots&\mathbf{A}^{s-1}\mathbf{x}(0)
\end{pmatrix}.
\end{split}
\end{align*}
{From  minimal realization theory ~\cite{Katayama05, Schutter00, Kailath80}, a system is {minimal} (that is, controllable and observable) if and only if}
\begin{align*}
\text{rank}(\mathcal{O}_r)=\text{rank}(\mathcal{C}_s)=\text{rank}(\mathbf{A})=|G|.
\end{align*}
From  Sylvester inequality~\cite{Horn13} we also have $\text{rank}(\mathbf{H}_{rs})=|G|$. Given that the rank of the Hankel matrix is equivalent to the model order~\cite{Kailath80}, {rank$(\mathbf{H}_{rs})=n(r,s\ge n)$}, we finally obtain:
\begin{equation}
|G|=\text{rank}(\mathbf{H}_{rs})=n.
\label{eq:rankHankel}
\end{equation}

Combining Eqs.~(\ref{eq:dimGN}) and (\ref{eq:rankHankel}), the model order $n$ is a strictly increasing function of the number of qubits $N$,
\begin{equation}
\frac{dn}{dN}>0,
\label{eq:nandN}
\end{equation}
and there exists  function $N=f(n)$ linking the two. 
Then, provided one can find a basis $B$ such that the {equivalent} classical state-space representation of the quantum model is minimal, we can link the quantum system dimension to a measurable quantity (the Hankel matrix rank). Although the requirement to find such a basis is a limitation of our method,  checking controllability and observability is straightforward. We do not even need to calculate the rank of the Hankel matrix, we can simply determine whether the system is {minimal} from the degree of the characteristic polynomial of the irreducible transfer function (See Eq.~(\ref{eq:degree}) and Appendices.~\ref{sec:example1}, \ref{sec:counterex}).}

Note that controllability and observability of the equivalent classical system is a sufficient but not necessary condition for Eq.~(\ref{eq:nandN}). Thus, it is sometimes possible (as shown in Appendix \ref{sec:counterex}) to perform dimension estimation even when these conditions do not hold.

We summarize the previous results in the following theorem, after introducing the concept of \textit{system space}, as the space given by $\mathcal S=(G_0, \mathcal{M}, \rho_0)$, that is, the three quantities that for a given basis uniquely determine the dynamics of the system and thus the accessible set $G$:
\begin{theorem} 
Given a system space $\mathcal S=(G_0, \mathcal{M}, \rho_0)$ expressed in a basis yielding a {minimal} realization {of the equivalent classical system},
and whose quantum dynamics generates correlation between the observed quantum probe and the rest of the qubits, 
the dimension of the quantum system $\mathcal{H}$ is given by:
\begin{align*}
\text{dim}(\mathcal{H})=2^{f_{\mathcal S}(n)},
\end{align*}
where $n=\text{rank}(\mathbf{H}_{rs}) ~(r,s\ge n)$ is the model order of the system and $f_{\mathcal S}$ is a strictly increasing function uniquely determined by the system space $\mathcal S$. 
\label{theorem}
\end{theorem}

To determine both $n$ and $f_\mathcal S$, we can  perform the following systematic procedure that estimates the dimension of an unknown quantum system.
\begin{enumerate}[leftmargin=3pt,itemindent=*, label={\bfseries Step \arabic*:}]
\item From the prior information about the interaction model and selecting a basis that gives a minimal realization, we theoretically obtain a recurrence relation between the model order and the quantum system dimension, yielding the function $f_{\mathcal S}(n)$.
\item We repeatedly prepare the initial state of the probe so that $\mathbf{x}(0)\neq \mathbf{0}$, and measure the probe at equally spaced  sampling points, obtaining the outcomes $y(j)=y(jdt)$. (see Sec.~\ref{sec:noisy} for details on how to choose the sampling time $dt$ from the sampling theorem).
\item We construct a sequence of square Hankel matrix $\{\mathbf{H}_{2,2}, \mathbf{H}_{3,3},\cdots, \mathbf{H}_{k,k},\cdots\}$, and calculate their determinant. In the absence of noise,  the dimension of the total system is given by the minimum $n$ such that the Hankel matrix determinant is zero:
\begin{align*}
\min_{n}\text{det}(\mathbf{H}_{n+1, n+1})=0.
\end{align*}  
\end{enumerate}
In the presence of noise, we can simply construct a large enough Hankel matrix and determine the corresponding $n$ by finding a singular value $\lambda_{n}$ such that $\lambda_n\gg \lambda_{n+1}$. (For details, see Sec. ~\ref{sec:noisy}).

 \subsection{Examples of Noiseless Dimension Estimation}  
Let us show a few examples of the dimension estimation procedure for  spin chain systems with nearest-neighbor couplings. Here, let the initial state of the quantum probe be the eigenstate of the observable which is measured so that $\mathbf{x}(0)\neq\mathbf{0}$. 
Also, we  assume that the rest of spins are in the maximally mixed state $\openone/2$, reflecting a practical scenario, where these spins are inaccessible and thus cannot be initialized.  This scenario is especially pertinent when we consider a system at room temperature: a paradigmatic example is the NV center spin, which can be used as a quantum probe to determine the Hamiltonian (and thus the structure) of other electronic and nuclear spins in a room-temperature molecule~\cite{Sushkov14l,Ajoy15}.

We first consider the following nearest-neighbor couplings without transverse fields: 
\begin{align*}
H_{\text{int}}=\sum_{k=1}^{N-1}\Big(\frac{A_k}{2} S^{\alpha}_kS^{\alpha}_{k+1}+\frac{B_k}{2} S^{\beta}_k S^{\beta}_{k+1}+\frac{C_k}{2} S^{\gamma}_k S^{\gamma}_{k+1}\Big),
\end{align*}
where $\{iS^{\alpha}_{k},iS^{\beta}_{k},iS^{\gamma}_{k}\}\in \mathfrak{su}(2)$ are the general spin-1/2 operators acting on $k$-th spin. Suppose that our quantum probe is coupled to the spin chain as a first spin. 

For the first example, let us consider the Heisenberg's model. For this case, $A_k\neq0, B_k\neq0$, but $C_k\neq0$. We can choose either $G_0=\{S^{\alpha}_1\}, \{S^{\beta}_1\}$ or $\{S^{\gamma}_1\}$ for our observable to measure. For this choice, the system is minimal. By induction, we can obtain the following relation: 
\begin{align*}
n=\begin{cases}
4^{N-1}-1~&(N:\text{odd})\\
4^{N-1}~&(N:\text{even})
\end{cases}.
\end{align*}
Therefore, the dimension of the total system for each case is given by:
\begin{equation}
\text{dim}(\mathcal{H})=
\begin{cases}
2\sqrt{n+1}~&(N:\text{odd})\\
2\sqrt{n}~&(N:\text{even})
\end{cases}.
\label{eq:Heisenberg0}
\end{equation}

As a second example, let us consider the exchange model. For this case, we can choose $A_k\neq0, B_k\neq0$ and $C_k=0$. Choosing $G_0=\{S^{\alpha}_1\}$ or $\{S^{\beta}_1\}$  enables us to obtain the smallest value of $n$. 
For this choice, the system is minimal. By induction, we can obtain the following relation~\cite{Sone17a}: $n=N$.
Therefore, the dimension of the total system is given by:
\begin{equation}
\text{dim}(\mathcal{H})=2^{n}.
\label{eq:XY0}
\end{equation}

As a third example, let us consider the Ising model. For this case, we can choose $A_k\neq0, B_k=C_k=0$. Choosing $G_0=\{S^{\beta}_{1}\}$ or $\{S^{\gamma}_1\}$, we  obtain $n=2$~\cite{Sone17a}. In this case the iteration Eq.~(\ref{eq:iterative}) saturates very quickly. Accessible set contains the correlation operator of the quantum probe and the spin next to the probe. 
This is an example of the case where only some spins become correlated to the probe; therefore, Theorem.~\ref{theorem} does not hold for this case. In order to obtain the dimension of the total system, we can employ the pulse sequence presented in~\cite{Sone17a} to obtain the exchange coupling in a good approximation.

Next, let us consider the case with transverse field, 
\begin{align*}
H=\sum_{k=1}^{N}\frac{\Omega_{k}}{2}S_{k}^{\gamma}+H_{\text{int}},
\end{align*}
where $\Omega_{k}\neq 0$. 
For the Heisenberg's model, by choosing $G_{0}=\{S_{1}^{\alpha}\}$ or $\{S^{\beta}_{1}\}$, {the system is  minimal} and we can obtain: $n=2^{2N-1}$. Therefore, the dimension of the total system is given by:
\begin{equation}
\text{dim}(\mathcal{H})=\sqrt{2n}.
\label{eq:Heisenberg1}
\end{equation}
For both the exchange model~($A_k\neq0, B_k\neq0$ and $C_k=0$) and Ising model~($A_k\neq0, B_k=C_k=0$), by choosing $G_0=\{S^{\alpha}_1\}$ or $\{S^{\beta}_1\}$, {the systems are minimal, and we can obtain: $n=2N$ \cite{Sone17a, Zhang14}. Therefore,  
\begin{equation}
\text{dim}(\mathcal{H})=2^{\frac{n}{2}}.
\label{eq:XYIsing1}
\end{equation}
Thus, in order to estimate the dimension of the Ising model, we can as well apply a transverse field to realize the global information propagation, so that all the spins become correlated to the quantum probe, instead of applying the control sequence presented in~\cite{Sone17a}.

\section{Noisy dimension estimation}
\label{sec:noisy}
Finally, let us consider how the  presence of background noise affects the estimation scheme. The measurement outcomes at each sampling time can then  be written as $y(j)+\zeta(j)$, where $\zeta(j)$ is the noise. The added noise usually leads the experimental Hankel matrix, of dimension $r \times r$, to have full rank,  $\text{rank}(\mathbf{H}_{r,r})=r$. It becomes then challenging to extract the exact model order $n$. 
We will show that it is still possible to obtain information about the ``true'' rank of the ideal Hankel matrix, by looking at the  singular values of the noisy matrix, at least for weak enough noise.

Let $\mathbf{H}_{r,r}$ be the $r\times r$ Hankel matrix obtained from the noisy experimental data. While we expect $\mathbf{H}_{r,r}$ to be full rank,  the true model order is $n<r$. 
Then, if the noise is weak, so that the experimental Hankel matrix is close enough to the ideal one, we expect that its singular values $\{\lambda_k\}_{k=1}^{r}$ in  decreasing order would be such that: $\lambda_1\ge \lambda_2\ge \cdots \lambda_n\gg \lambda_{n+1}\ge\cdots \lambda_{r}>0$. Indeed, although $\{\lambda_{l}\}_{l=n+1}^{r}$ are non-zero due to the noise, they should still be small, provided  the noise is comparatively small. Therefore, by constructing a sequence of singular value ratios $\{\lambda_{k}/\lambda_{k+1}\}_{k=1}^{r-1}$, we can expect to be able to determine the most probable model order $n$  by finding a sharp peak at $\lambda_{n}/\lambda_{n+1}$. 

We investigate the robustness of this method by simulating the nearest-neighbor exchange model without transverse field $H=\sum_{k=1}^{N-1}\frac{J_k}{2}(S^x_k S^x_{k+1}+S^y_k S^y_{k+1})$ for $N=4,5,6$ qubits.
 For this Hamiltonian, the true model order is $n=N$. The quantum probe is the first qubit, and we estimate the model order by measuring only $X_1$. From  the sampling theorem, the sampling time $dt$ can be chosen as $dt=\pi/\omega_{\text{max}}$, where $\omega_{\max}$ is the largest eigenvalue of the Hamiltonian. We note that in principle, $\omega_{\max}$ is unknown. An upper bound can be determined only by assuming a maximum system dimension and a maximum parameter strength (and then assuming all the coupling take this maximum value). 
 
 We consider an added noise $\zeta(j)$  characterized by a normal distribution with variance $10^{-7}-10^{-4}$, which corresponds to $35-20~[\text{dB}]$. The results of these simulations are shown in Figs.~\ref{fig:4qubit}-\ref{fig:6qubit}.

\begin{figure}[t!]
\includegraphics[width=1\linewidth]{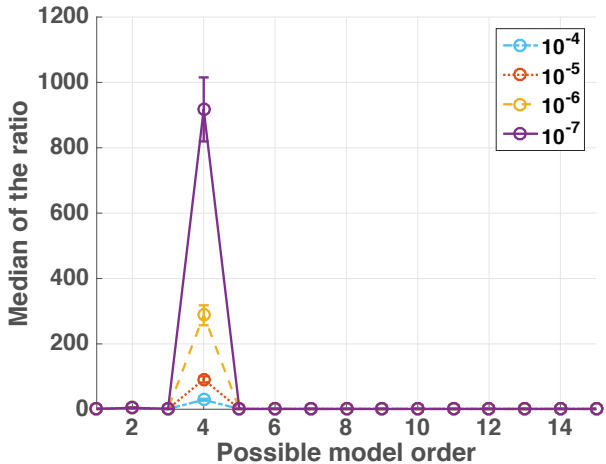}
\caption{$\bf{N=4}$: Actual model order is $n=4$.  We plot the median of the ratios between the singular values $\{\lambda_{k}/\lambda_{k+1}\}_{k=1}^{15}$ over 500 random Hamiltonian realizations as a function of the possible model order $k$. We choose the sampling time $d t$  by assuming that the number of qubits $N$ could be at most $10$ and all the coupling strengths  take the possible maximum number $100$ (this assumptions set $\omega_{\text{max}}$). For each Hamiltonian, we construct $100\times 100$ Hankel matrix $\mathbf{H}_{100}$ and estimate $\{\lambda_{k}/\lambda_{k+1}\}$ 100 times to evaluate the average. Solid line with circles: $\sigma^2=10^{-7}$; dashed line with circles: variance $\sigma^2=10^{-6}$; dotted line: variance $\sigma^2=10^{-5}$; dashed-dotted line: variance $\sigma^2=10^{-4}$. The error bars are the median of the standard deviation of the singular value ratio over 500 random Hamiltonian realizations.
The final sharp peak occurs at $k=4$, and from $k=5$ the ratios are almost the same. The final peak becomes sharper as the variance of the noise becomes  smaller. }
\label{fig:4qubit}
\end{figure}

\begin{figure}[t!]
\includegraphics[width=1\linewidth]{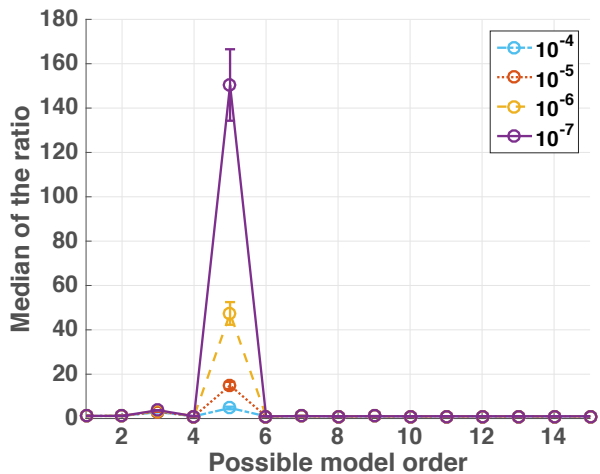}
\caption{$\bf{N=5}$: Actual model order is $n=5$. Simulation details are the same as in Fig.~\ref{fig:4qubit}. The final sharp peak occurs at $k=5$, and from $k=6$ the ratios are almost same. The final peak becomes sharper as the variance of the noise becomes  smaller. }
\label{fig:5qubit}
\end{figure}

\begin{figure}[th!]
\includegraphics[width=1\linewidth]{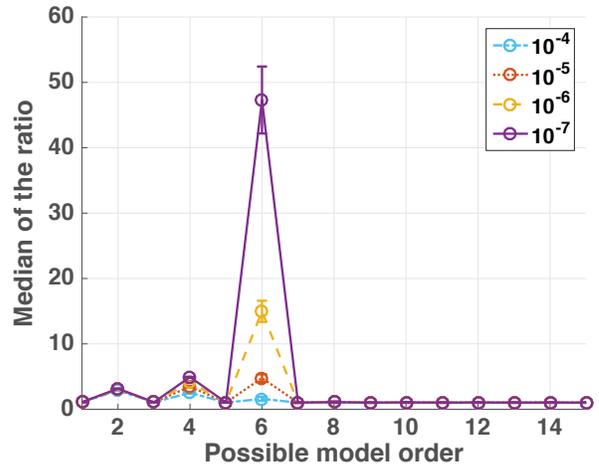}
\caption{$\bf{N=6}$: Actual model order is $n=6$. Simulation details are the same as in Fig.~\ref{fig:4qubit}. We can see that when SNR~$\ge 25\text{dB}$ the final peak occurs at $k=6$, and we can see the flatting part from $k=7$. The final peak becomes much sharper as the variance of the noise becomes smaller. }
\label{fig:6qubit}
\end{figure}

We can conclude that when the Signal-to-Noise Ratio (SNR) is large enough, the final peak tends to occur at the true model order. Also, the weaker the noise, the sharper the peak  at the true model order. We find, however, that when the number of qubits increases, the peak becomes less sharp. This means that we need larger SNR to estimate the dimension of the total system. From these results, we can expect that from $N=7$, we need at least SNR~$\ge 35\text{dB}$, which is demanding to realize in current laboratory conditions. Therefore, we can conclude that this method can be useful only for a few-body interacting system. {Also note that the method is model-sensitive, which means that the function that we have to use to estimate the dimension of the system is strictly sensitive to the interaction model. Thus, we need to use the modified function for the suspected model to find the dimension. For example, if we allow some noise on $XY$ Hamiltonian, for example, $C_k\ll 1$, in this case, we need to employ the function for the Heisenberg's model.}

\section{Conclusions} 
\label{sec:conc}
We have theoretically studied the problem of estimating the dimension of an interacting qubit system. We assumed that the target system model is given as prior information, but that the system can be accessed only indirectly via a quantum probe. These assumptions considerably relax the required experimental resources, as well as other implicit assumptions about the controllability and observability of the target system. 
Provided the coupling model allows the generation of non-local correlations with the quantum probe during the system dynamics, the dimension of the system can be exactly estimated from sampling the quantum probe evolution at discrete time steps. The estimation is based on measuring the rank of the Hankel matrix constructed from the experimental data, a step in the recently proposed procedure for the similar problem of Hamiltonian identification~\cite{Zhang14,Zhang15,Sone17a}. 

This study provides a useful application of local quantum probes, which can serve as a quantum sensor to determine the exact dimension of the system. Our method can also be employed  more broadly to determine the dimension of  a general interacting $N-$body qudit system, i.e. a system with dimension $\text{dim}(\mathcal{H})=d^N$: as long as $d$ and the interaction model $\mathcal{M}$ are given,  conservation of the operator algebraic structure ensures that a similar procedure than described here can be applied.

 In the presence of noise, which changes the rank of the experimentally obtained Hankel matrix, we numerically show that the dimension of the system can still be estimated by finding the final peak in the sequence of ratios of the Hankel matrix singular values. We conclude that the methodology has an acceptable performance for weak noise and small number of qubits. Also, the method is model-sensitive because we must employ the modified function for the suspected model to find the dimension if we allow some noise on the model itself.


As an outlook, we remark that the same method could be applied to the exact dimension estimation of an open quantum systems. Based on the application of ERA in the open quantum system identification, as proposed in~\cite{Zhang15}, we can also determine the dimension with Markovian dynamics if the model of the coupling and decoherence dynamics are both given as prior information. 

\begin{acknowledgements}
We would like to offer our gratitude to Can Gokler, Quntao Zhuang, Shuichi Adachi, Yuanlong Wang, and Mohan Sarovar
 for fruitful discussions. 
This work was supported in part by the U.S. Army Research Office through Grants No. W911NF-11-1-0400 and W911NF-15-1-0548 and by the NSF PHY0551153. 
\end{acknowledgements}

\appendix

\section{Example: Finding the model order $n$ for the Ising model with transverse field}
\label{sec:example1}

An important step in dimension estimation is to find the function $N=f(n)$ that links the system dimension to the model order $n$. Here we provide an explicit example on how to find this function for the Ising model with transverse field. The procedure for other models would be analogous. 

Specifically, here we want to derive Eq.~(\ref{eq:XYIsing1}), using a proof by induction. Let us consider the following Hamiltonian:
\begin{align*}
H=\sum_{k=1}^{N}\frac{\Omega_{k}}{2}S^{z}_{k}+\sum_{k=1}^{N-1}\frac{J_{k}}{2}S^{x}_{k}S^{x}_{k+1}.
\end{align*}
The operator set for this model is thus $\Gamma^{(N)}=\{S^{z}_1,~...~,~S^{z}_{N}, ~S^{x}_{1}S^{x}_2,~...~,~S^{x}_{N-1}S^{x}_{N}\}$. Let us choose $G_{0}=\{S_{1}^{x}\}$ as the observable set. The interaction model is described by the sequence of graphs satisfying the following adjacency matrices:
\begin{align*}
[\Xi(\{\mathcal{G}_{k\ge 1}\})]_{i,j}=
\begin{cases}
1,~\text{when $i=j\pm1$ and $i=j$}\\
0,~\text{otherwise}.
\end{cases}
\end{align*}
with the coupling type of nearest-neighbor Ising coupling with single qubit energies.

In order to make sure that $\langle S_{1}^{x}(0)\rangle\neq0$, we can choose:
\begin{align*}
\rho_{0}=\frac{1}{2}
\begin{pmatrix}
1&1\\
1&1
\end{pmatrix}
\otimes
\frac{1}{2^{N-1}}\openone^{\otimes (N-1)}.
\end{align*}

For any system dimension (that is, any number $N$ of qubits), we can perform the iterative procedure of Eq.~(\ref{eq:iterative}). Given the chosen observable set,  $G_0=\{S_1^x\}$, $[\![G_0,~\Gamma^{(N)}]\!]=\{S_1^y\}$ so that 
\begin{align*}
G_1=G_0\cup[\![G_0,~\Gamma^{(N)}]\!]=\{S_1^x, ~S_1^y\}.
\end{align*} 
Since $S_1^y$ does not commute with the Hamiltonian, it will yield further contributions under the operation
$[\![S_1^y,\Gamma^{(N)}]\!]=\{S_1^x,~ S_1^zS_2^x\}$. We thus obtain
\begin{align*}
G_2=\{S_1^x,~S_1^y,~S_1^zS_2^x\}.
\end{align*} 
Similarly, we can calculate $G_3=G_2\cup [\![G_2,\Gamma^{(N)}]\!]$, which contains new contributions arising from $S_1^zS_2^x$ :
\begin{align*}
G_3=\{S_1^x,~S_1^y,~S_1^zS_2^x,~S_1^zS_2^y\}.
\end{align*}
This iterative procedure saturates after {$(2N-1)$} steps, as no new operators are generated. We then obtain the following accessible set:
\begin{equation}
\begin{split}
G^{(N)}=\{S^x_1, ~&S^y_1, ~S^z_1 S^x_2, ~S^z_1 S^y_2,~...~,\\
&S^z_1\cdots S^z_{N-1}S^x_N, ~S^z_1\cdots S^z_{N-1}S^y_N\},
\end{split}
\label{eq:IsingAccessibleSet}
\end{equation} 
as it can be proved by induction:
\begin{enumerate}
\item{$N=2$: By the iterative procedure in Eq.~(\ref{eq:iterative}),  the accessible set can be explicitly found to be:
\begin{align*}
G^{(2)}=\{S^x_1, ~S^y_1, ~S^z_1 S^x_2, ~S^z_1 S^y_2\},
\end{align*}
which satisfies Eq.~(\ref{eq:IsingAccessibleSet}).
}
\item{$N=w+1 \ (\forall w\in \mathbb{N})$: Let us assume that for $w$ qubits we obtained
\begin{align*}
\begin{split}
G^{(w)}=\{S^x_1, ~&S^y_1, ~S^z_1 S^x_2, ~S^z_1 S^y_2,~...~,\\
&S^z_1\cdots S^z_{w-1}S^x_w, ~S^z_1\cdots S^z_{w-1}S^y_w\}.
\end{split}
\end{align*} 
}
\end{enumerate}
Then, for  $N=w+1$, the Hamiltonian operator set is:
\begin{align*}
\Gamma^{(w+1)}=\Gamma^{(w)}\cup\{S^{z}_{w+1}, ~S^{x}_{w}S^{x}_{w+1}\}
\end{align*}
and by the iterative procedure in Eq.~(\ref{eq:iterative}) we  obtain:
\begin{align*}
G^{(w+1)}=G^{(w)}\cup\{S^{z}_1\cdots S^{z}_{w}S^{x}_{w+1}, ~S^{z}\cdots S^{z}_{w}S^{y}_{w+1}\},
\end{align*}
which yields:
\begin{align*}
\begin{split}
G^{(w+1)}=\{S^x_1, ~&S^y_1, ~S^z_1 S^x_2, ~S^z_1 S^y_2,~...~,\\
&S^z_1\cdots S^z_{w}S^x_{w+1}, ~S^z_1\cdots S^z_{w}S^y_{w+1}\}.
\end{split}
\end{align*}
This demonstrates that  Eq.~(\ref{eq:IsingAccessibleSet}) also holds for $N=w+1$. Therefore,  we can conclude that for the Ising model with transverse field, Eq.~(\ref{eq:IsingAccessibleSet}) holds, yielding
\begin{align*}
|G^{(N)}|=2N.
\end{align*}

{By constructing a state-space representation, the system matrix $\tilde{\mathbf{A}}$ becomes a $2N\times 2N$ skew-symmetric matrix with the only nonzero elements $\tilde{\mathbf{A}}_{2k,2k-1}=\Omega_k$ and $\tilde{\mathbf{A}}_{2k+1, 2k}=J_k$. Since we want to measure $G_0=\{S_1^{x}\}$, the output matrix is $\mathbf{C}=\begin{pmatrix}1 & 0&\cdots&0\end{pmatrix}\in\mathbb{R}^{2N}$. Given the initial state, the initial coherent vector is $\mathbf{x}(0)=(1,~ 0,~\cdots,~0)^{T}\in\mathbb{R}^{2N\times 1}$, and $(\tilde{\mathbf{A}}, \mathbf{C}, \mathbf{x}(0))$ generates an irreducible transfer function $T(s)=\mathbf{C}(s\mathbf{I}-\tilde{\mathbf{A}})^{-1}\mathbf{x}(0)=P(s)/Q(s)$, where $\text{deg}(Q(s))=2N$. Note that as we mentioned in the Sec.~\ref{sec:intro}, the model order is the number of poles of the irreducible transfer function; therefore, given an irreducible transfer function $T(s)=P(s)/Q(s)$, we have always $n=\text{deg}(Q(s))$. From Eq.~(\ref{eq:rankHankel}), when the system is {minimal}, we also have $\text{deg}(Q(s))=|G|$. In this way, we can check {observability and controllability of the system} by finding the degree of the transfer function's characteristic polynomial.

{For the Ising model with transverse field, since $|G|=\text{deg}(Q(s))=2N$}, the system is {minimal}, and the model order is $n=2N$, that is, the system dimension is given by
\begin{align*}
\text{dim}(\mathcal{H})=2^{N}=2^{n/2}.
\end{align*}
Note that for the other models presented in Sec.~\ref{sec:noiseless}, a similar proof can be applied.

\section{{Example of dimension estimation without controllability or observability}}
\label{sec:counterex}
Here, let us show an example to show that $\frac{dn}{dN}>0$ cannot ensure that the system is {minimal}. This can be understood in the following. While the operators describing full correlation between the rest of qubits and the single quantum probe can be directly generated, but the choice of the initial states and the measurement observables can also change the controllability and observability of the system. 
 An example is associated with the following Hamiltonian:
\begin{align*}
H=\sum_{k=1}^{N-1}\Big(\frac{J_{k}}{2}S^{z}_{k}S^{z}_{k+1}+\frac{L_{k}}{2}S^{z}_{k}S^{x}_{k+1}\Big).
\end{align*}
Suppose that we choose $G_{0}=\{S^{x}_{1}\}$ or $G_{0}=\{S^{y}_{1}\}$. Actually, the model orders for these different choices of observable sets are the same. Let us choose $G_{0}=\{S^{x}_{1}\}$. Then, from the iterative procedure in Eq.~(\ref{eq:iterative}), we can obtain: 
\begin{align*}
\begin{split}
G=\{S^{x}_1,~&S^{y}_1 S^{z}_2,~S^{y}_1 S^{x}_2, ~...,~\\
&S^{y}_1\cdots S^{y}_{N-1}S^{z}_{N}, S^{y}_1\cdots S^{y}_{N-1}S^{x}_{N}\},
\end{split}
\end{align*}
 which contains all the directly generated operators describing the correlations of all qubits with a single quantum probe, and  
 \begin{align*}
 |G|=2N-1.
 \end{align*}
By constructing a state-space representation, $\tilde{\mathbf{A}}$ becomes a $(2N-1)\times (2N-1)$ skew-symmetric matrix with the only nonzero elements $\tilde{\mathbf{A}}_{2k, 2k-1}=J_{k}$ and $\tilde{\mathbf{A}}_{2k+1, 2k-1}=L_{k}$. Since we want to measure $G_0=\{S^{x}_{1}\}$, our output matrix should be $\mathbf{C}=\begin{pmatrix}1&0&\cdots&0\end{pmatrix}\in\mathbb{R}^{1\times (N-1)}$. From the initial sate, our initial coherent vector is $\mathbf{x}(0)=(1, 0, \cdots, 0)^T\in\mathbb{R}^{(2N-1)\times 1}$. Then, we can find that the irreducible transfer function $T(s)=P(s)/Q(s)$ has the property of  $\text{deg}(Q(s))=2N-2$, which means that the model order is 
\begin{align*}
n=|G|-1=2N-2,
\end{align*} 
and 
\begin{align*}
\frac{dn}{dN}>0
\end{align*}
so that $n$ is an increasing function of $N$. In this case, the system dimension becomes: 
\begin{align*}
\text{dim}(\mathcal{H})=2^{\frac{N+1}{2}}.
\end{align*}
However, we have $n\neq|G|$, which indicates that the system cannot be {minimal}. This shows that there exist systems either not controllable or observable, even if we can directly generate the operators describing the correlation between all the qubits and a single quantum probe. In this particular case, the system is both 
not controllable and not observable because 
\begin{align*}
\text{rank}(\mathcal{O}_{r})=\text{rank}(\mathcal{C}_{s})=|G|-1,
\end{align*}
which means that there is an extra operator generated, which is not needed to be counted in order to describe the dynamics of the system, and the system can be reduced into the one with the effective size. Therefore, the controllability and observability are strictly dependent on the choice of the observable set $G_0$ and the initial state $\rho_0$.

\bibliography{Biblio}

\end{document}